\documentclass[11pt,twoside]{article}
\usepackage{asp2010}

\resetcounters

\begin{document}

\title{Radial velocity monitoring of long period hot subdwarf +\\
main sequence binaries with HERMES@Mercator}
\author{Roy~H.~{\O}stensen and Hans~Van Winckel}
\affil{Instituut voor Sterrenkunde, K.U.~Leuven,
%      Celestijnenlaan 200D,
B-3001 Leuven, Belgium}

\begin{abstract}
Population synthesis models predict that the majority of sdBs form through
stable mass transfer leading to long period binaries.
To date, about a hundred close short-period binaries with an sdB component
have been found, but not a single long-period system has been established.
We show preliminary results from our recent effort in determining
orbits for a sample of long-period sdB systems using the {\sc hermes}
spectrograph on the Mercator telescope.
\end{abstract}

\section{Introduction}
Binarity is a key component both in the formation and in the final
evolution of stars, when expansion forces one of the companions to
fill its gravitational well, and its envelope to spill over to that
of the partner. Binary processes strongly influence the composition
of stellar populations, and a diverse variety of evolved objects
can only be explained through the direct interaction between the bloated
envelope of a giant, and accretion by its companion.
Such binary interactions are complex and the models invoke a number
of poorly constrained parameters,
such as the efficiency of envelope ejection, the physical description of the
common-envelope phase, the accretion efficiency on to the companion,
and many others.
Observations that can constrain these model parameters are
essential not just for explaining the particular systems, but for
population synthesis in general, and for extrapolating to environments
of various densities and metalicities as found in clusters and other galaxies.

\begin{table}[t]
\caption{Eight targets in our sample of composite hot subdwarf / main sequence
systems. The spectroscopic class from various sources is given together
with the visual magnitude and the number spectra we have collected so far.
}\smallskip
\begin{center}\small
\begin{tabular}{llrrl}\tableline\noalign{\smallskip}
Target name & Sp.Class & $m_V$ & N.sp & References \\
\noalign{\smallskip}\tableline\noalign{\smallskip}
BD--11$^\circ$162 & sdO+? & 11.2 & 12 & \citet{zwicky57} \\
                  & sdO+G &      &    & \citet{berger80} \\
                  & sdOB+K0 &    &    & \citet{ulla98} \\
PG 1104+243       & sdB+K2 & 11.3& 24 & \citet{ferguson84} \\
                  & sdB+G8 &     &    & \citet{orosz97} \\
Balloon 82800003  & sdB+K1 & 11.4& 16 & \citet{bixler91} \\
BD+29$^\circ$3070 & sdOB+F & 10.4& 15 & \citet{berger80} \\
                  & sdB+K0 &     &    & \citet{ulla98} \\
BD+34$^\circ$1543 & sdB+F  &  9.4& 13 & \citet{berger80} \\
                  & sdB+G8 &     &    & \citet{ulla98} \\
BD--7$^\circ$5977 & sdB+K0IV-III & 10.5 & 17 & \citet{ulla98} \\
                  &        &     &    & \citet{heber02} \\
Feige 80          & sdO+A  & 11.4& 21 & \citet{berger80} \\
                  & sdO+G8 &     &    & \citet{ulla98} \\
Feige 87          & sdB+?  & 11.7& 15 & \citet{jeffery98} \\
\noalign{\smallskip}\tableline
\end{tabular} \end{center} \end{table}

\citet{han02,han03} made a thorough binary population synthesis study
of the hot subdwarfs, using all three binary formation channels that are
thought to contribute significantly to the population. The three are;
(1) If the subdwarf progenitor has a low mass companion, then mass transfer
on the RGB is unstable, and the orbit will shrink until the envelope is
ejected.  The study of \citet{maxted01} as recently completed by
\citet{copperwheat11} finds that $\sim$50\%\ of all
sdB stars reside in short-period binary systems ($P_{\mathrm{orb}} < 10$\,d).
(2) If the companion is more massive than the subdwarf (at least at the
end of mass transfer), the orbit will have expanded substantially.
Such orbits are hard to measure, but the companion
can be detected spectroscopically or from infra-red excess.
\citet{napiwotzki04} found that more than a third of their sdB sample
show the spectroscopic signature of main sequence (MS) companions, while
\citet{reed04}, using 2MASS photometry, inferred that about half
of the sdBs in the field have main-sequence companions, and are therefore
likely to be of this post-stable-Roche-lobe-overflow (pRLOF) type.
(3) The final binary formation channel is the merger of two low-mass white
dwarfs, and has a much lower efficiency than the two other channels.

By now, $\sim$100 sdB stars are known to reside in short period binaries, and
they are providing clear constraints for common-envelope ejection models.
A recent compilation can be found in Appendix A of \citet{geier11a}, but new 
systems are discovered at a high pace. The longest period systems are 15 and
29 days respectively \citep[from][]{morales-rueda03}.
For the longer period systems very little is known. \citet{green01} mention
a mean $\Delta v\sin i$ of 11.5\,km/s from 89 observations of 19 composite
binaries, and estimate periods averaging 3--4 years, but provide no
details about particular systems.
Here we will.

\begin{figure*}[t]
\centering
\includegraphics[width=\textwidth]{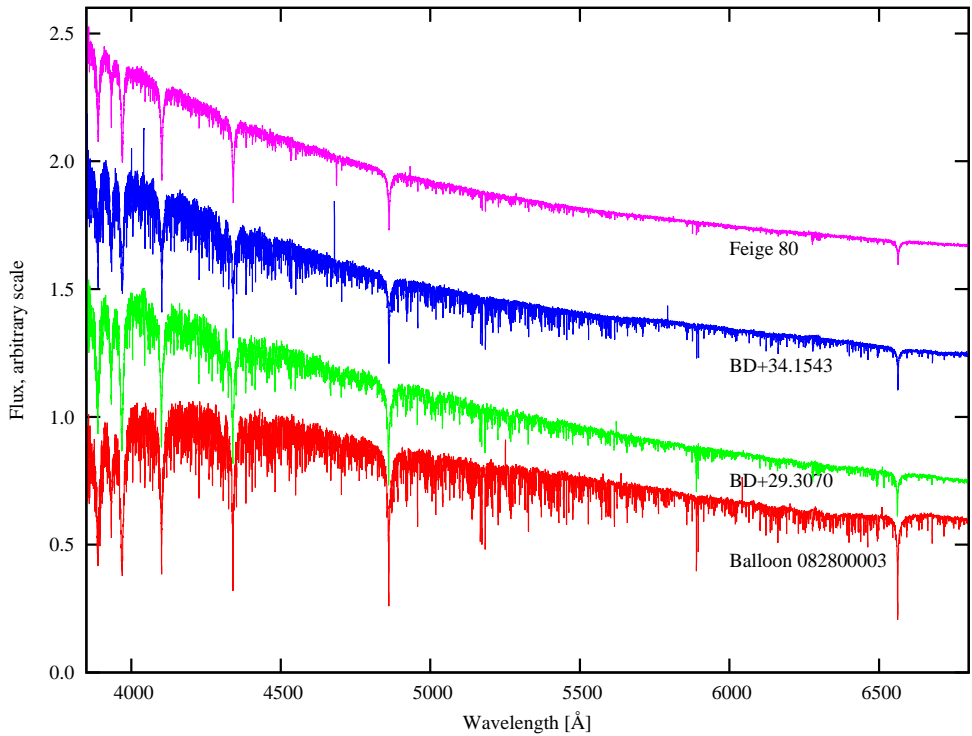}
\caption{Single spectra for four of the targets in the sample.
The resolution has been substantially degraded, and the wavelength
region truncated to 3800--6800\,\AA\ for illustration purposes.
The slope of each spectrum has been corrected by calibration with
observations of the single sdB star Feige\,66, and offset for
clarity. It can be seen that
the contribution from the companion varies substantially between
the targets. However, even in Feige\,80, where the spectrum is
well dominated by the sdB component, there are sufficient
lines from the MS companion for the cross-correlation to
succeed in establishing precise radial velocity measurements
for the cool star.  }
\end{figure*}

\section{Observations}
The observations presented here were all made with the Mercator telescope
on La Palma, which is a twin of the Swiss 1.2m Euler telescope at La Silla.
In November 2008 a new state-of-the-art fibre-fed ultra-stable high-resolution
Echelle spectrograph was installed. This instrument, dubbed
{\sc hermes} \citep[an acronym for {\em High Efficiency and Resolution Mercator
Echelle Spectrograph,}][]{raskin11}
reaches a spectral resolution, $R$\,=\,85\,000 over a spectral
range covering 3770 to 9000\,\AA\ in a single exposure, and has a peak
efficiency of 28\%. Being mounted in a temperature and pressure controlled
environment provides the stability to ensure reliable velocity determinations
over extended periods of time.
A substantial fraction of the observing time on {\sc mercator} is 
dedicated to a long-term program to establish orbits of evolved
binary systems.
On this program we have, since the commissioning of {\sc hermes},
made regular observations of a sample of composite hot subdwarf stars. 
The high resolution and excellent time coverage provided by
{\sc hermes} will allow us to establish the periods and velocity
amplitudes of even the longest period systems with very high precision.

\begin{figure*}[t]
\centering
\includegraphics[width=\textwidth,clip]{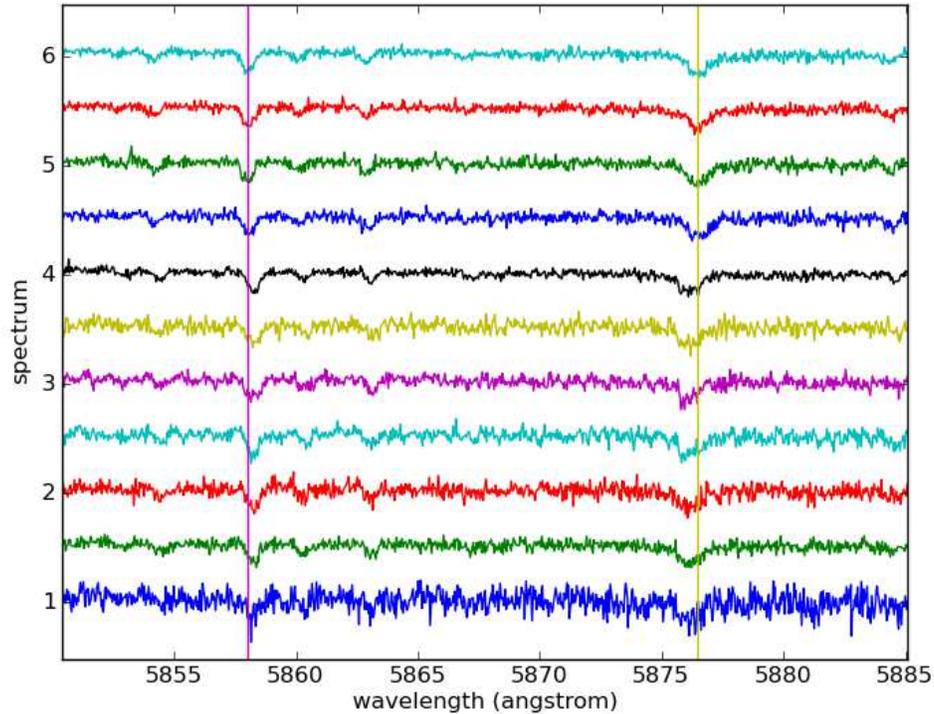}
\caption{A small section of our spectra of BD+34$^\circ$1543.
The He\,{\sc i} line from the sdB at 5876\,\AA\ is clearly seen to
move in the opposite phase to that of the Ca\,{\sc i} line at
5857\,\AA.}
\end{figure*}

In Table~1 we list eight stars in our sample that we have determined preliminary
orbits for. The full sample contains 22 stars, but some of these have been
discontinued since they were too faint to determine reliable radial
velocities, and others have been inserted to replace those, but do not
have a sufficiently long time base yet.
The limiting magnitude for {\sc hermes} to
reach an S/N useful for our method appears to be around $V$\,=\,13, but
we have focused on the stars brighter than $V$\,=\,12\ in order to use
the available time most efficiently.
In the table we also list the spectroscopic class from the literature and
the reference for that identification. The $V$ magnitude of the stars
are also listed together with the number of spectra available for each object.
This is the total number of spectra, and in some cases they are taken on
the same night so that the number of points useful for period determination
can be smaller.

In Figure~1 we show spectra of four of the stars up to 6800\,\AA.
The top spectrum is Feige\,80 (=\,PG\,1317+123), and it is 
clearly the star where the hot component gives the strongest
contribution to the total flux. The He\,{\sc ii} line at 4686\,\AA\ is
strong and deep while He\,{\sc i} at 4472 is quite shallow and hard to
discern, making this component an sdO star, as noted already by \citet{berger80}.
With so little detectable He\,{\sc i}, the temperature must be close to
~50\,000\,K, and is therefore likely to be in the post-EHB stage of evolution.
The lines from the companion, although weak, are indicative of an
F--G companion, but we have not attempted to make a reliable determination
of the companion classes for any of the sample objects yet. 
The exception on the other extreme is BD--7$^\circ$5977, which is completely
dominated by the deep and narrow spectral lines of a K subgiant or early RGB
star (spectrum not shown).

Another clear sdO star in our sample is BD--11$^\circ$162, which shows no
trace of the He\,{\sc i} 4472 line, making it even hotter than Feige\,80,
however, the companion is also stronger contributing about 1/3 of the
flux around 6000\,\AA.
The other objects in Figure~1 are more typical of the systems in our sample,
with both stars contributing roughly equally to the flux around 6000\,\AA.
Balloon 82800003, BD+34$^\circ$1543, BD+29$^\circ$3070 and Feige\,87 all
show the He\,{\sc i} lines
at 4472 and 5876\,\AA, but no detectable He\,{\sc ii}, making them sdB stars.
PG\,1104+243 shows both He\,{\sc i} and He\,{\sc ii} lines, making it an
sdOB star. The classifications are summarised in Table~2, together with results
from the orbital analysis.

Inspection of the high-resolution spectra clearly reveals large differences
in the rotational broadening of the lines from the cool companion.
BD+29$^\circ$3070 shows lines broadened by as much as 60\,km/s, while most
of the stars have moderately broadened lines. The subgiant component of
BD--7$^\circ$5977 has very sharp lines, consistent with no significant
rotation, as one may expect from an expanding giant.

\begin{figure*}[t]
\centering
\includegraphics[width=0.3\textwidth,angle=-90,clip]{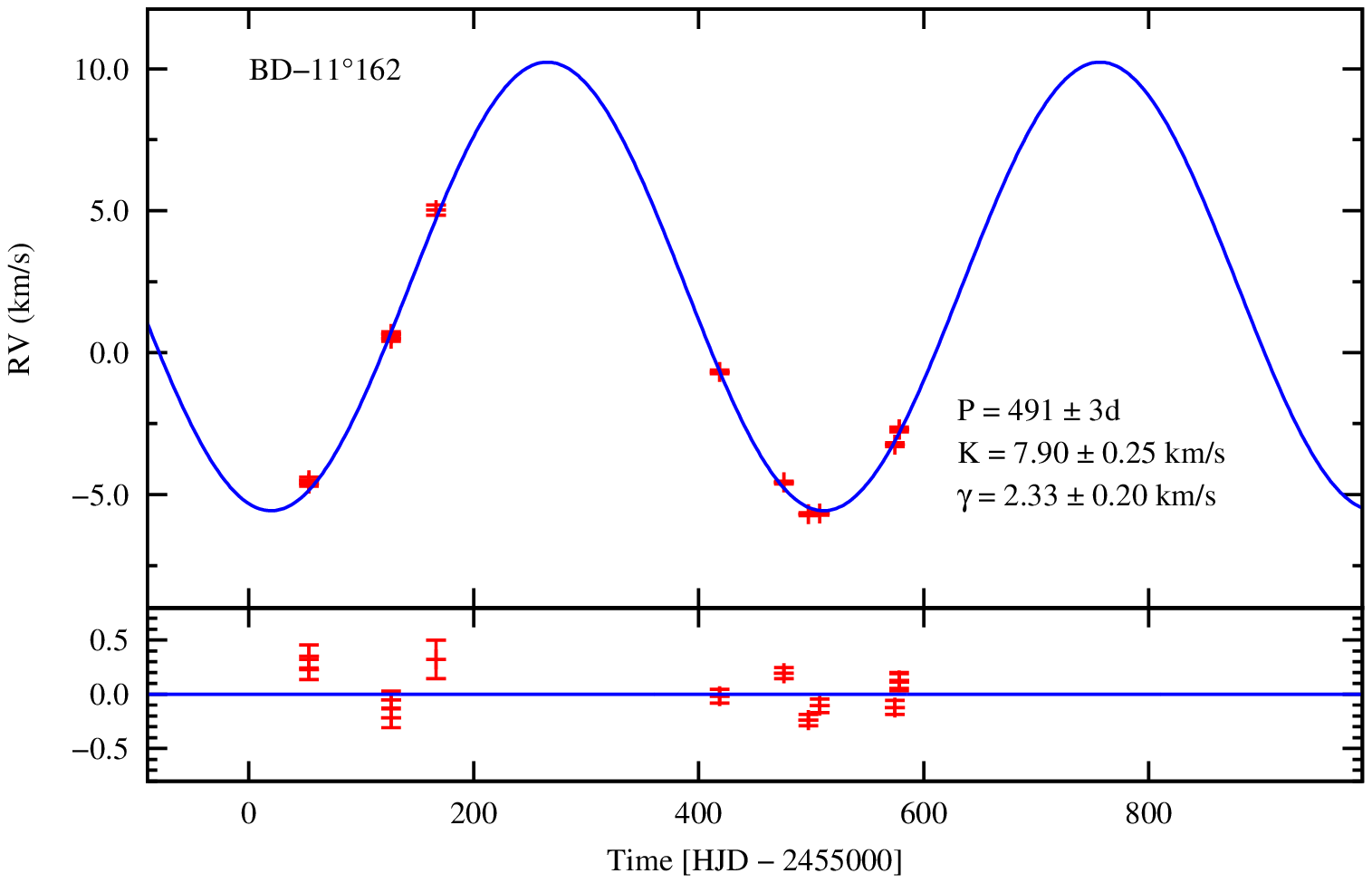}
\includegraphics[width=0.3\textwidth,angle=-90,clip]{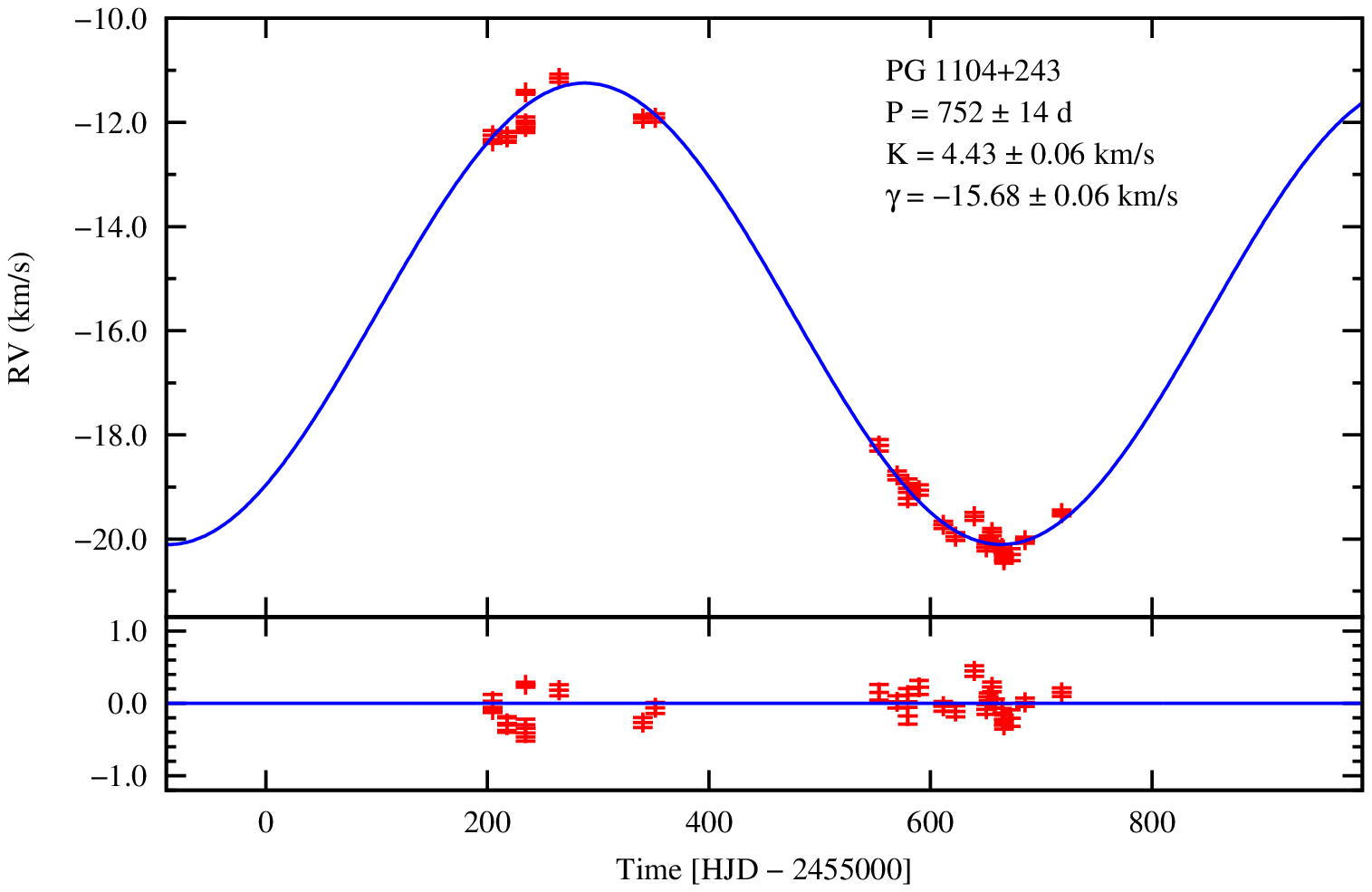}\\
\includegraphics[width=0.3\textwidth,angle=-90,clip]{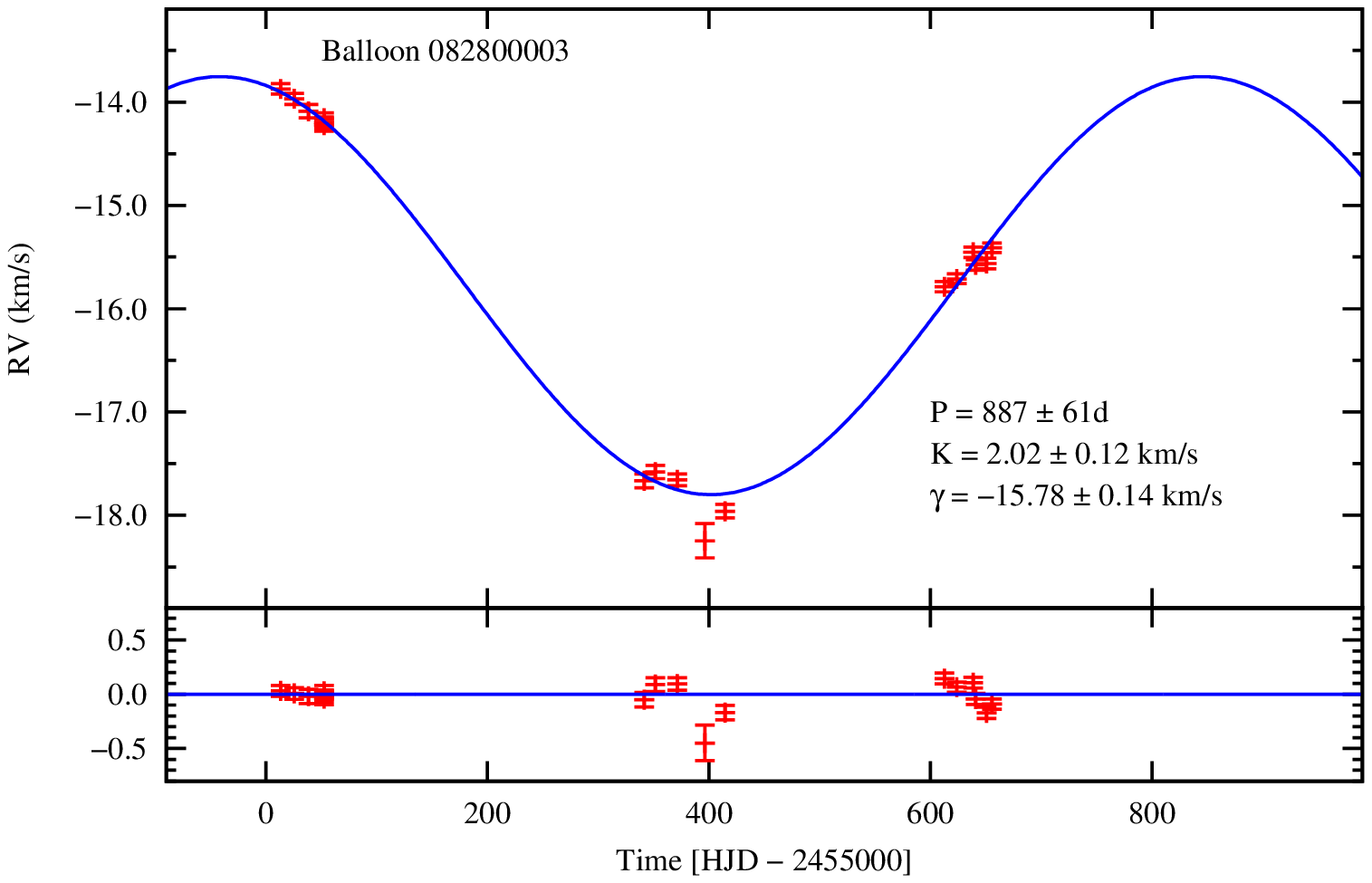}
\includegraphics[width=0.3\textwidth,angle=-90,clip]{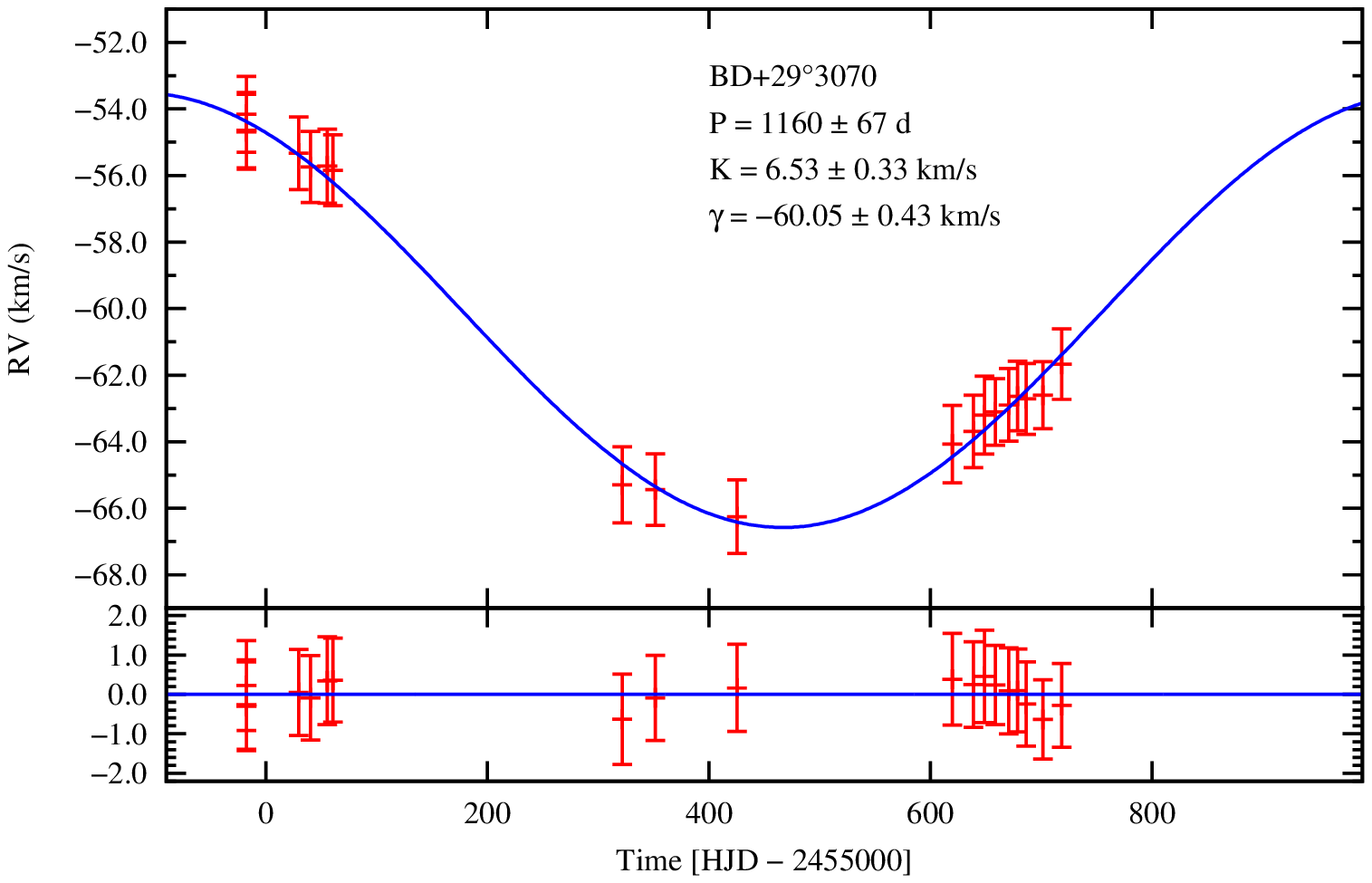}\\
\caption{
Thanks to the high precision of the {\sc hermes} spectrograph,
the twelve observations on nine different nights
of BD--11$^\circ$162 are sufficient to derive a
quite reliable period after just two observing seasons, as long as
circular orbits are assumed.
Other stars yield similarly excellent results.
For BD+29$^\circ$3970 the cross correlation procedure overestimates
the errors since the high rotation makes the line profiles substantially
non-Gaussian.  }
\end{figure*}

\section{Radial velocity analysis}

In Figure~2 we show a small section of the spectra that make out
the sequence of observations for BD+34$^\circ$1543.
On the left side one can see the  Ca\,{\sc i}
line at 5857\,\AA\ from the cool MS star
(as well as the two weaker components of the triplet),
and on the right the broader He\,{\sc i} line at 5876\,\AA\ originating
from the sdB star can be seen.
The two lines are clearly moving in antiphase with each other,
and the sdB would appear to have an amplitude about twice that of the
F-star, as one would expect.

Up to now, we have not made any attempt at deriving radial velocities
from the lines of the hot subdwarf, although in principle it should be
possible for most of the targets. The results we show here are all
based on cross-correlation analysis of the lines from the main sequence
companion with standard templates. Due to the high number of lines
available this procedure yields excellent high precision radial velocities
even when the spectra are relatively low S/N. In Figure~3 and 4 we show
the measurements and fits for the eight stars listed in Table~1.
As one can see from the plots, all objects have rather poor sampling,
and only the shortest case covers more than a full orbit, the rest showing
only between half and one cycle. There is however exceptionally little scatter
around the fitted sine curves, which clearly indicates that the assumption
of circular orbits as expected from binary evolution with stable mass transfer
is valid.

A few of the stars, Feige\,80\ in particular, shows a scatter around the
fitted curves that are significantly higher than the cross-correlation
errors indicates. We believe that these variations are real, as they have
the amplitude and periods expected for gamma Doradus type pulsations
in F-stars or spots in main-sequence stars. Several examples of low amplitude
photometric variability in sdB+F/G stars, consistent with these scenarios,
were detected in {\em Kepler} targets \citep{ostensen11b}. For the particular
case of BD+29$^\circ$3070 the opposite is the case; the cross-correlation
errors are clearly larger than expected when compared with the sine fit.
This is due to the discrepancy between the rotationally broadened
line profiles, and the Gaussian profiles assumed by the cross-correlation
procedure.

The first object shown in Figure~3 is BD--11$^\circ$162.
With the errors from the cross-correlation routine being less than
0.1 km/s for all the the measurements, the fit to the RV curve is excellent.
With 12 observations covering more than 500 days, it appears that
we have just managed to cover a complete orbit in two observing seasons.
Our solution provides an orbital period of 491 days with
an RV amplitude of 7.9 km/s, as stated on the plot. This is the
shortest period found in any of the stars. The next object in
Figure~3, PG\,1104+243, gives the second longest period (~750\,d).
Four of the remaining objects come out with periods that are between
818 and 915 days, while two objects stand out with periods between
1100 and 1200 days. For all the stars the radial velocity amplitudes
are low, between 2 and 8 km/s.  Our best fit periods and velocity
amplitudes are given in Table~2.

\begin{figure*}[t]
\centering
\includegraphics[width=0.3\textwidth,angle=-90,clip]{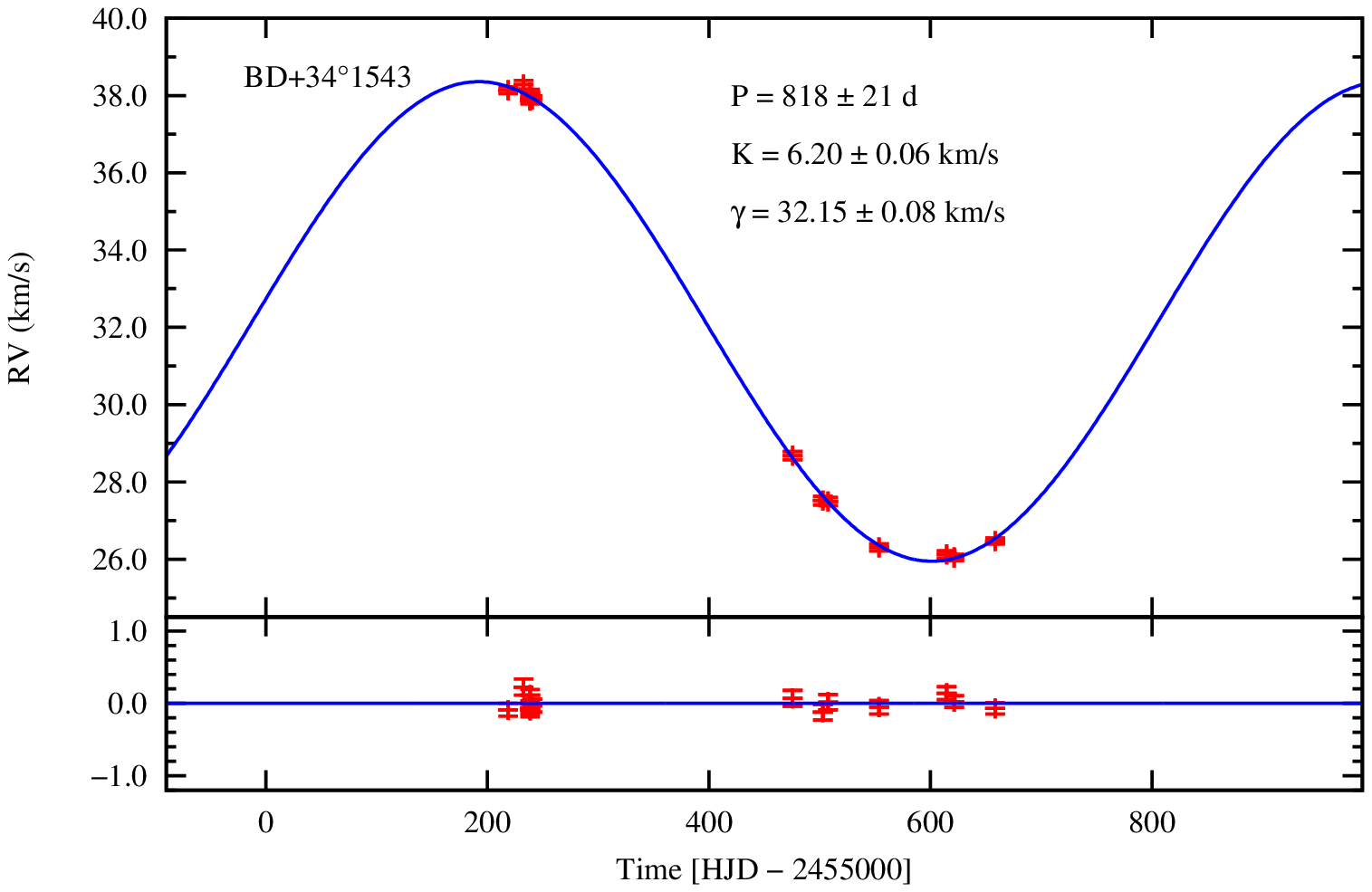}
\includegraphics[width=0.3\textwidth,angle=-90,clip]{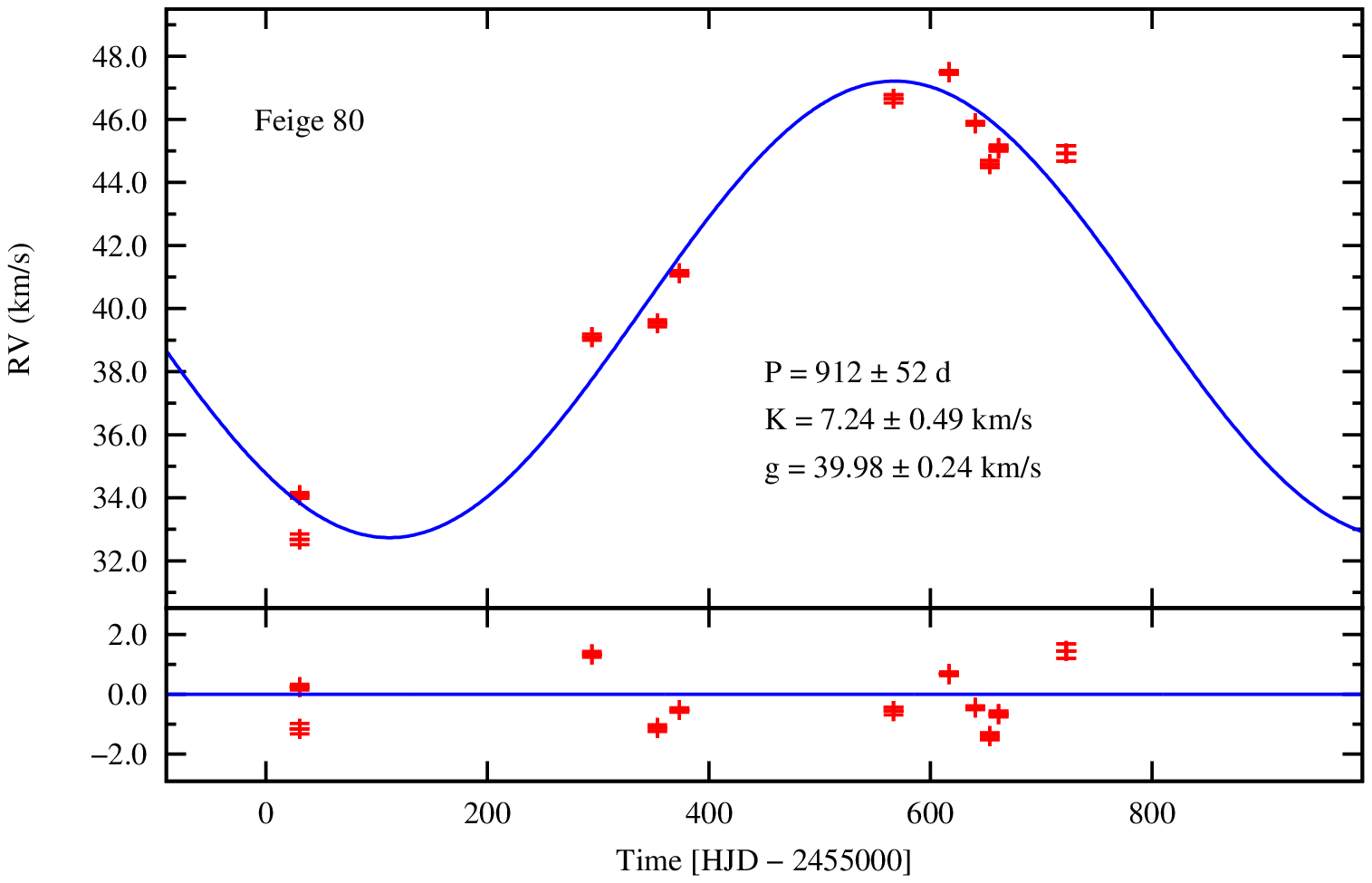}\\
\includegraphics[width=0.3\textwidth,angle=-90,clip]{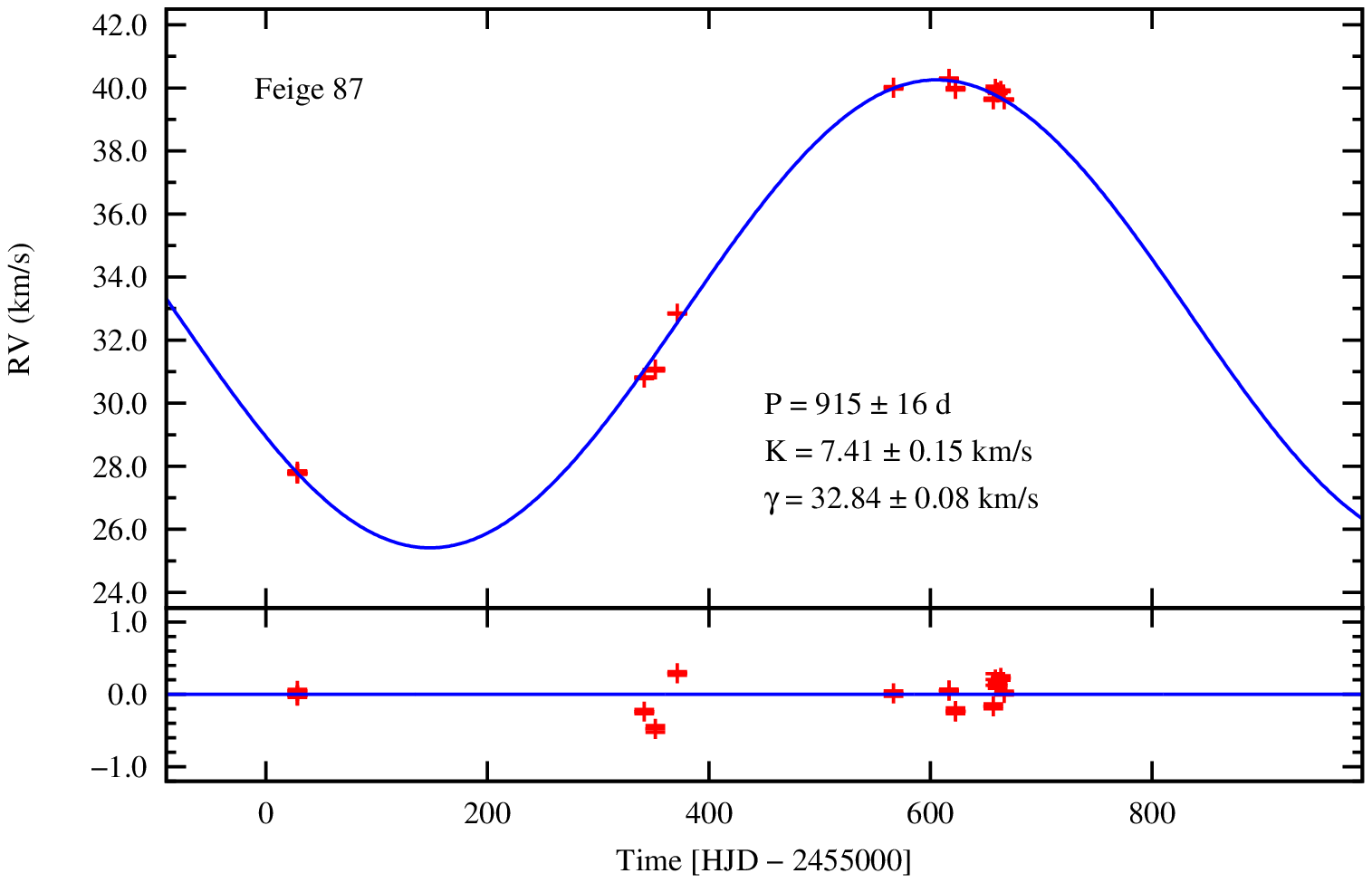}
\includegraphics[width=0.3\textwidth,angle=-90,clip]{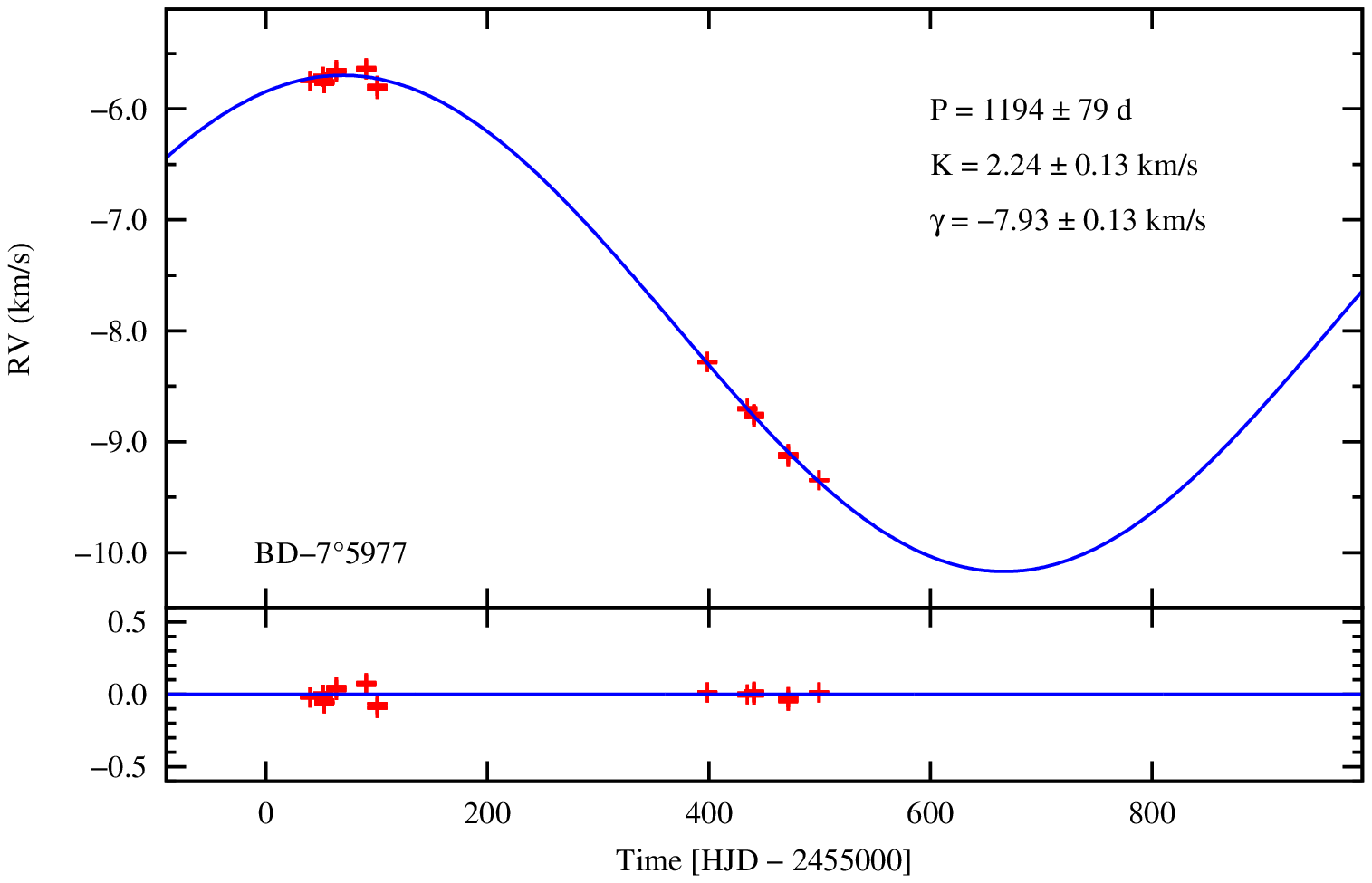}\\
\caption{
Same as Figure~2, for four other stars that we were able to find reasonable
solutions for.
For Feige\,80 the residuals are significantly larger than indicated
by the error bars. We have not detected anything amiss with the spectroscopy
so we believe that these are real velocity variations originating
on the surface of the F-star, most likely caused by stellar pulsations.
}
\end{figure*}

\begin{table}[t]
\caption{Summary of orbital periods, velocity amplitudes for the main
sequence companion and system velocities found for the sample stars.
The classes given are our own estimates based on the
appearance of the spectra. In the last column the rotational broadening
is given. The numbers in parentheses are the estimated errors on the
values. For the period and velocity amplitude the stated errors are
the formal fitting errors from non-linear least square three-parameter
sine fits.
The rotational broadening was not fitted, but estimated by convolving
a model spectrum until a reasonable agreement was achieved.
}\smallskip
\begin{center}\small
\begin{tabular}{llrrrrr}\tableline\noalign{\smallskip}
Target name & Sp.Class & $m_V$ & Period
            & $K_{\mathrm{MS}}$ & $\gamma$ & $v_{\mathrm{rot}}$sin($i$) \\
\noalign{\smallskip}\tableline\noalign{\smallskip}
BD--11$^\circ$162&sdO+G    &11.2& 491(3) & 7.9(3) &+2.3(2) & 5(5) \\
PG 1104+243      &sdOB+G   &11.3& 752(14)& 4.43(6)&--15.68(6) & 5(5) \\
Balloon 82800003 &sdB+F    &11.4& 887(61)& 2.0(1) &+15.8(2) & 20(5) \\
BD+29$^\circ$3070&sdOB+F   &10.4& 1160(67)& 6.5(3) &--60.1(4) & 60(5) \\
BD+34$^\circ$1543&sdB+F    & 9.4& 818(21)& 6.20(6)&+32.15(8)&12(3) \\
BD--7$^\circ$5977&sdB+K2III&10.5& 1194(79)& 2.2(1) &--7.9(1)& 0(1) \\
Feige 80         &sdO+G    &11.4& 912(52)& 7.2(5) &+40.0(2)& 15(5) \\
Feige 87         &sdB+G    &11.7& 915(16)& 7.4(2) &+32.8(1)&  5(5) \\
\noalign{\smallskip}\tableline
\end{tabular} \end{center} \end{table}

\section{Conclusions}
Our long time-base spectroscopy from the Mercator telescope has revealed
orbits of $\sim$500\,d and longer for eight sdB+F/G binaries (Figure~2 and 3).
While pRLOF systems are expected to be found with a wide range of periods,
a strong peak is predicted to be found just above 100\,d
\citep[see Figure~21 of][]{han03}.
All the systems shown in Figure~1 appear to have periods much longer
than this peak, indicating that some of the parameters that govern
the mass transfer process need to be adjusted.

While more than one hundred hot subdwarfs are known to exist in short-period
systems, the long-period systems predicted as the outcome of stable Roche-lobe
overflow on the first giant branch require extraordinary efforts to nail
down.
We have presented our first results from a dedicated survey to reveal
the orbital periods of hot subdwarfs with main sequence companions, and the
periods detected so far are far longer than the distribution predicted
by \citet{han03}, but in excellent agreement with the estimated average
of 3--4 years mentioned by \citet{green01}.

\acknowledgements
The research leading to these results has received funding from the European
Research Council under the European Community's Seventh Framework Programme
(FP7/2007--2013)/ERC grant agreement N$^{\underline{\mathrm o}}$\,227224
({\sc prosperity}), as well as from the Research Council of K.U.Leuven grant
agreement GOA/2008/04.

Based on observations obtained with the {\sc hermes} spectrograph, which is supported
by the Fund for Scientific Research of Flanders (FWO), Belgium,
the Research Council of K.U.Leuven, Belgium,
the Fonds National Recherches Scientific (FNRS), Belgium,
the Royal Observatory of Belgium, the Observatoire de Gen\'eve,
Switzerland and the Thüringer Landessternwarte Tautenburg, Germany.

\bibliographystyle{asp2010}
\bibliography{sdbrefs}

\end{document}